\documentclass[prl,superscriptaddress,floatfix,twocolumn]{revtex4}
\usepackage{amsfonts}
\usepackage{stmaryrd}
\usepackage{bbm}
\usepackage{mathrsfs}
\usepackage{tipa}
\usepackage{amssymb}
\usepackage{txfonts}
\usepackage{graphicx}
\usepackage{dcolumn}
\usepackage{epstopdf}
\usepackage{array}
\usepackage{color}

\newcommand{\PreserveBackslash}[1]{\let\temp=\\#1\let\\=\temp}
\newcolumntype{C}[1]{>{\PreserveBackslash\centering}p{#1}}
\newcolumntype{R}[1]{>{\PreserveBackslash\raggedleft}p{#1}}
\newcolumntype{L}[1]{>{\PreserveBackslash\raggedright}p{#1}}

\begin{document}

\newcommand*{\cm}{cm$^{-1}$\,}

\title{\textcolor{black}{Charge-Density-Wave-Induced Peak-Dip-Hump Structure and the Multiband Superconductivity in a Kagome Superconductor CsV$_{3}$Sb$_{5}$}}

\author{Rui Lou}
\thanks{lourui@lzu.edu.cn}
\affiliation{School of Physical Science and Technology, Lanzhou University, Lanzhou 730000, China}
\affiliation{Leibniz Institute for Solid State and Materials Research, IFW Dresden, 01069 Dresden, Germany}
\affiliation{Helmholtz-Zentrum Berlin f{\"u}r Materialien und Energie, Elektronenspeicherring BESSY II, 12489 Berlin, Germany}

\author{Alexander Fedorov}
\thanks{a.fedorov@ifw-dresden.de}
\affiliation{Leibniz Institute for Solid State and Materials Research, IFW Dresden, 01069 Dresden, Germany}
\affiliation{Helmholtz-Zentrum Berlin f{\"u}r Materialien und Energie, Elektronenspeicherring BESSY II, 12489 Berlin, Germany}

\author{Qiangwei Yin}
\affiliation{Department of Physics and Beijing Key Laboratory of Opto-electronic Functional Materials $\textsl{\&}$ Micro-nano Devices, Renmin University of China, Beijing 100872, China}

\author{Andrii Kuibarov}
\affiliation{Leibniz Institute for Solid State and Materials Research, IFW Dresden, 01069 Dresden, Germany}

\author{Zhijun Tu}
\affiliation{Department of Physics and Beijing Key Laboratory of Opto-electronic Functional Materials $\textsl{\&}$ Micro-nano Devices, Renmin University of China, Beijing 100872, China}

\author{Chunsheng Gong}
\affiliation{Department of Physics and Beijing Key Laboratory of Opto-electronic Functional Materials $\textsl{\&}$ Micro-nano Devices, Renmin University of China, Beijing 100872, China}

\author{Eike F. Schwier}
\affiliation{Experimentelle Physik VII, Universit{\"a}t W{\"u}rzburg, Am Hubland, 97074 W{\"u}rzburg, Germany}
\affiliation{W{\"u}rzburg-Dresden Cluster of Excellence ct.qmat, Germany}

\author{Bernd B{\"u}chner}
\affiliation{Leibniz Institute for Solid State and Materials Research, IFW Dresden, 01069 Dresden, Germany}
\affiliation{Institute for Solid State and Materials Physics, TU Dresden, 01062 Dresden, Germany}

\author{Hechang Lei}
\thanks{hlei@ruc.edu.cn}
\affiliation{Department of Physics and Beijing Key Laboratory of Opto-electronic Functional Materials $\textsl{\&}$ Micro-nano Devices, Renmin University of China, Beijing 100872, China}

\author{Sergey Borisenko}
\thanks{s.borisenko@ifw-dresden.de}
\affiliation{Leibniz Institute for Solid State and Materials Research, IFW Dresden, 01069 Dresden, Germany}

\begin{abstract}
  The entanglement of charge density wave (CDW), superconductivity, and topologically nontrivial electronic structure has recently been discovered
  in the kagome metal $A$V$_3$Sb$_5$ ($A$ = K, Rb, Cs) family. With high-resolution angle-resolved photoemission spectroscopy, we study the electronic properties of \textcolor{black}{CDW and superconductivity in CsV$_3$Sb$_5$}. The spectra around $\bar{K}$ is found to exhibit a peak-dip-hump structure associated with two separate branches of dispersion, demonstrating the isotropic CDW gap opening below $E_{\rm F}$. The peak-dip-hump lineshape is contributed by linearly dispersive Dirac bands in the lower branch and a dispersionless flat band close to $E_{\rm F}$ in the upper branch.
  \textcolor{black}{The electronic instability via Fermi surface nesting could play a role in determining these CDW-related features.
  The superconducting gap of $\sim$0.4 meV is observed on both the electron band around $\bar{\Gamma}$ and the flat band around $\bar{K}$, implying the
  multiband superconductivity. The finite density of states (DOS) at $E_{\rm F}$ in the CDW phase are most likely in favor of the emergence of multiband superconductivity, particularly the enhanced DOS associated with the flat band. Our results not only shed light on the controversial origin of the CDW, but also offer insights into the relationship between CDW and superconductivity.}
\end{abstract}

\maketitle

Recently, exploring the mutual interactions of different correlated phases, magnetism, and non-trivial band topology became of great scientific
interest, where novel quantum phenomena such as quantum anomalous Hall effect \cite{DengY2020,SerlinM2020,LiuC2016}, topological superconductivity \cite{ZhangP2018,WangD2018,ChenC2019,ZhuS2020,XuG2016}, pair density wave \cite{RuanW2018,AgterbergD2020}, and topological Chern magnetism \cite{YinJX2020,WangQ2021} have been intensively studied. A peak-dip-hump lineshape induced by symmetry breaking or electron-boson coupling usually arises in the single-particle spectrum of the correlated systems like
actinides \cite{DasT2012}, cuprates \cite{KordyukA2002,DamascelliA2003}, iron chalcogenides \cite{LiuZK2013}, and colossal magnetoresistive manganites \cite{MannellaN2005,MannellaN2007}. Comprehensive knowledge of the renormalized electronic states and the redistribution of spectral weight inside the peak-dip-hump structure is essential for uncovering the physics of intertwined correlated states.

The kagome metal $A$V$_3$Sb$_5$ ($A$ = K, Rb, Cs) with a non-zero $Z_2$
topological invariant has recently been discovered to host charge density wave (CDW) and superconductivity, arousing numerous interests in the community \cite{OrtizB2019,OrtizB2020,YangSY2020,YinQ2021,OrtizB2021,JiangY2020,YuFH2021,ZhaoCC2021,ZhaoH2021,DuanW2021,ChenX2021,ZhangZ2021,LiangZ2021,LiHX2021,
ChenH2021,NiS2021,ZhouX2021,XiangY2021,MuC2021,FuY2021,SongD2021,ZhuC2021,XuH2021,FengXL2021,DennerM2021}. Scanning tunneling microscopy (STM) \cite{LiangZ2021} and x-ray scattering \cite{LiHX2021} measurements have demonstrated a three-dimensional charge modulation with a 2 $\times$ 2 $\times$ 2 superstructure,
which associates with the CDW formation at 80--110 K \cite{OrtizB2019,OrtizB2020,OrtizB2021,YinQ2021,FuY2021}.
\textcolor{black}{The softening of acoustic phonon is absent near the CDW vector \cite{LiHX2021}, and no CDW amplitude mode is
identified in a time-resolved spectroscopy experiment \cite{WangZX2021}, suggesting the CDW may be unconventional. On the other hand, an electronic Raman scattering measurement reports the existence of in-plane phonon anomalies below $T_{\rm CDW}$ \cite{WulferdingD2021}, and an optical spectroscopy
study suggests that the Fermi surface (FS) nesting is important in the CDW formation \cite{ZhouX2021}, supporting the theoretical
calculations \cite{TanHX2021}. A STM study on KV$_3$Sb$_5$ reports that the charge modulation exhibits an intensity reversal in real space across the CDW gap, as expected in the Peierls CDW scenario \cite{JiangY2020}. Overall, the origin of CDW in $A$V$_3$Sb$_5$ is still controversial.
The intricate interplay of CDW and superconductivity is revealed in CsV$_3$Sb$_5$ under external hydrostatic pressure and uniaxial strain \cite{YuFH2021NC,ChenKY2021,QianTM2021}. Further studies of the relationship between these two phases can facilitate uncovering the paring mechanism of
superconductivity.}

Several angle-resolved photoemission spectroscopy (ARPES) studies were carried out \cite{NakayamaK2021,WangZG2021,KangM2021,HuY2021,LiuZH2021,ChoS2021},
of which many efforts have been devoted to resolving the CDW gap in CsV$_3$Sb$_5$. Nevertheless, two helium-lamp-based measurements report strong anisotropy of CDW gap opening around $\bar{K}$ \cite{NakayamaK2021,WangZG2021}, in sharp contrast to the nearly isotropic gap behavior around $\bar{K}$ from another synchrotron-based study \cite{KangM2021}. \textcolor{black}{In order to clarify the discrepancy, and gain insights into the origin of CDW as well as its relation to superconductivity, further ARPES measurements are desired.}

\textcolor{black}{In this work, we study the CDW and superconducting states of CsV$_3$Sb$_5$ using ARPES. We find the isotropic CDW gap opening involved in
a peak-dip-hump structure around $\bar{K}$, where the peak structure exhibits a flat band close to $E_{\rm F}$. The FS nesting could play a role in the
formation of CDW order. The superconducting gap of $\sim$0.4 meV is observed over multiple FSs around $\bar{\Gamma}$ and $\bar{K}$. We suggest that the finite density of states (DOS) at $E_{\rm F}$ in the CDW phase are most likely in favor of the multiband superconductivity, particularly the enhanced DOS associated with the flat band.}

\begin{figure}[t]
  \setlength{\abovecaptionskip}{-0.5cm}
  \setlength{\belowcaptionskip}{-0.2cm}
  \begin{center}
  \includegraphics[trim = 4.5mm 0mm 0mm 0mm, clip=true, width=1.05\columnwidth]{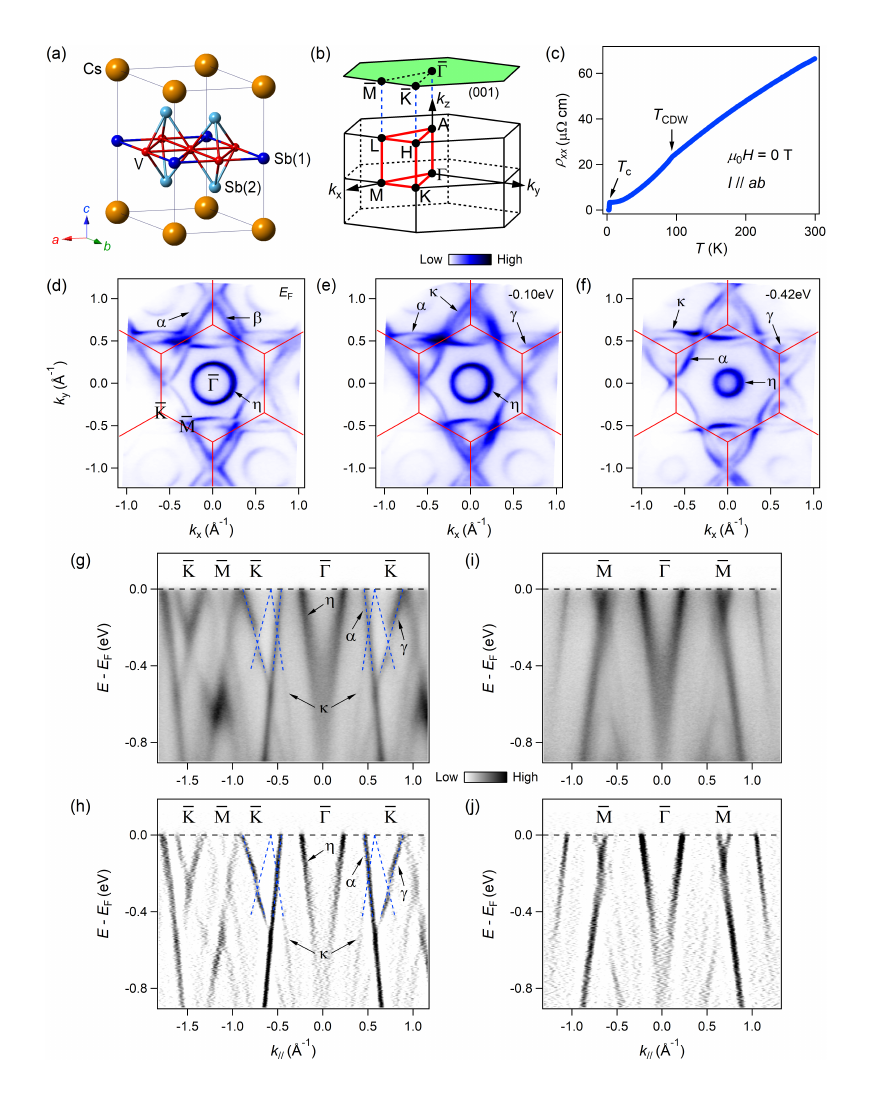}
  \end{center}
  \caption{
  (a) Schematic crystal structure of CsV$_3$Sb$_5$ in the pristine phase.
  (b) Bulk and (001)-projected BZs in the pristine phase.
  (c) Temperature dependence of the resistivity $\rho_{xx}$ at $\mu_0H$ = 0 T.
  (d)-(f) Constant-energy ARPES intensity plots ($T$ = 1.4 K, $h{\nu}$ = 95 eV) at binding energies of 0, -0.10, and -0.42 eV, respectively.
  The red solid lines indicate the pristine BZs.
  \textcolor{black}{(g),(h) Photoemission intensity plot and corresponding second derivative plot ($T$ = 1.4 K) along the $\bar{\Gamma}$-$\bar{K}$-$\bar{M}$ direction taken with 95-eV photons, respectively. The blue dashed lines indicate the linearly dispersive bands around $\bar{K}$.
  (i),(j) Same as (g),(h) recorded along the $\bar{\Gamma}$-$\bar{M}$ direction.}
  }
\end{figure}

High-quality single crystals of CsV$_3$Sb$_5$ were grown by the self-flux method \cite{FuY2021,SM}. As shown in Fig. 1(a), the pristine phase of CsV$_3$Sb$_5$ crystallizes in a layered hexagonal structure
with space group $P$6/$mmm$ (No. 191) \cite{OrtizB2019,OrtizB2020,OrtizB2021,YinQ2021,FuY2021}, consisting of an alternative stacking of Cs
layer and V-Sb slab along the $c$ axis, \textcolor{black}{between which the cleavage is anticipated (see Fig. S1 for the evidence of nonpolar surface).}
The two-dimensional (2D) kagome lattice is formed by V atoms coordinated by Sb atoms, which are categorized into in-plane Sb(1) and out-of-plane Sb(2) sites.
The schematic bulk and (001)-projected Brillouin zones (BZs) of pristine phase are presented in Fig. 1(b).
According to previous ARPES studies, the electronic structure of $A$V$_3$Sb$_5$ shows quasi-2D nature \cite{HuY2021,LiuZH2021}, we hereafter use the denotation of high-symmetry points in 2D BZ. In Fig. 1(c), the zero-field
resistivity exhibits metallic behavior above the superconducting transition temperature ($T_{\rm c}$ $\sim$ 2.5 K) except a kink
at $\sim$92 K, indicating the onset of CDW transition \cite{FuY2021}.

\begin{figure}[b]
  \setlength{\abovecaptionskip}{-0.15cm}
  \setlength{\belowcaptionskip}{-0.2cm}
  \begin{center}
  \includegraphics[trim = 4.8mm 0mm 0mm 0mm, clip=true, width=1.04\columnwidth]{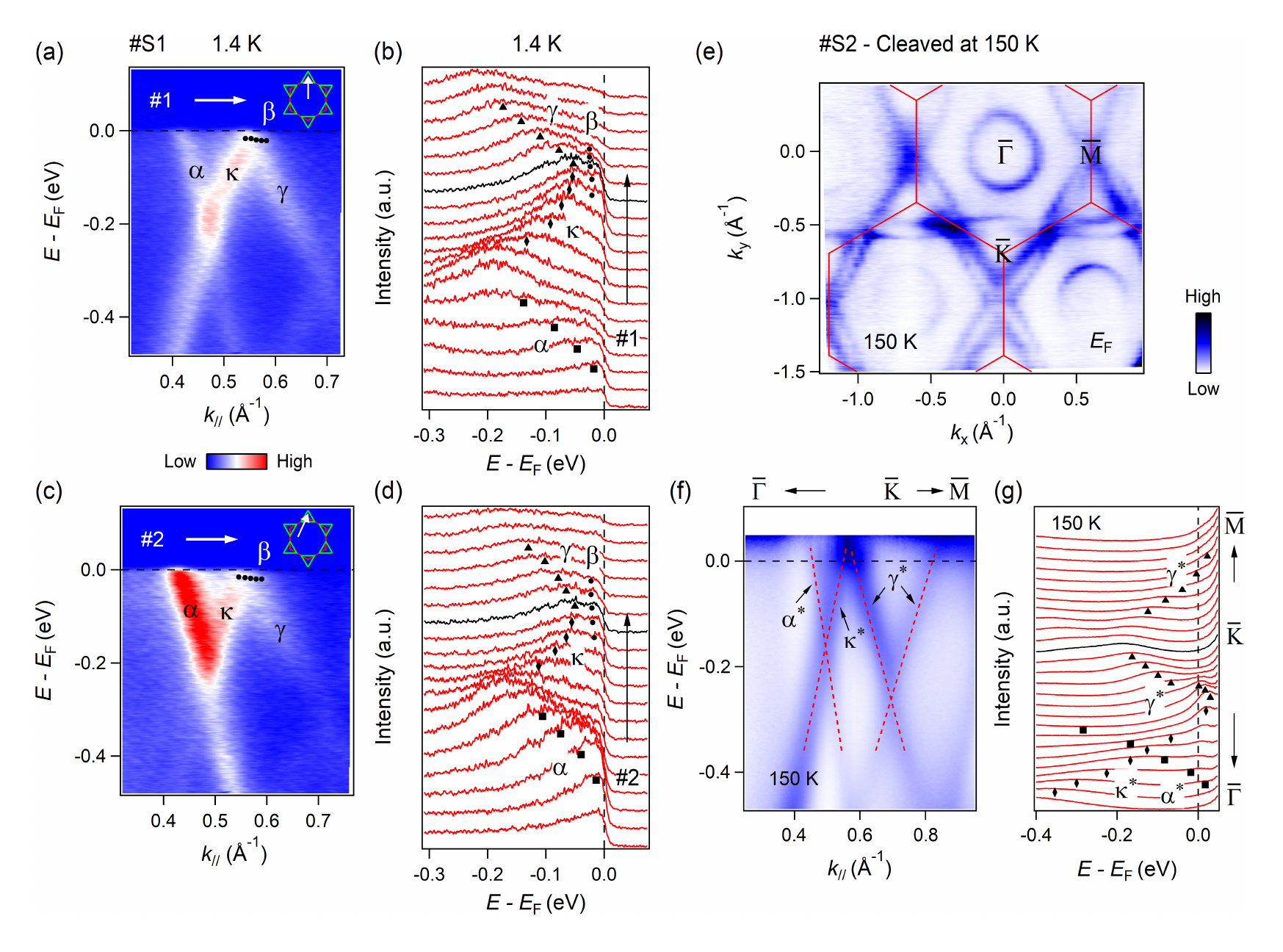}
  \end{center}
  \caption{
  \textcolor{black}{(a) ARPES intensity plot ($T$ = 1.4 K, $h\nu$ = 50 eV, sample \#S1) measured along cut \#1, which is illustrated by the white arrow in the
  inset. The black dots of $\beta$ are determined by the EDCs in (b).
  Inset: Sketches of the $\beta$ FS around $\bar{K}$ with pristine BZ.
  (b) EDC plot of (a) over the momentum range of 0.40 $\textless$ $k_{\varparallel}$ $\textless$ 0.66 $\AA^{-1}$.
  (c),(d) Same as (a),(b) recorded along cut \#2, which is illustrated by the white arrow in the inset of (c).
  (e) FS mapping of a sample cleaved at 150 K ($T$ = 150 K, $h\nu$ = 95 eV, sample \#S2).
  (f) ARPES intensity plot ($T$ = 150 K) measured along the $\bar{\Gamma}$-$\bar{K}$-$\bar{M}$ direction with 50-eV photons. The image is divided by the Fermi-Dirac distribution function convolved with the instrumental energy resolution. The red dashed lines are guides to the eye for the band dispersions.
  (g) EDC plot of (f) over the momentum range of 0.40 $\textless$ $k_{\varparallel}$ $\textless$ 0.90 $\AA^{-1}$.}
  }
\end{figure}

The constant-energy mappings at binding energies of 0, -0.10, and -0.42 eV are displayed in Figs. 1(d)-1(f), respectively.
Focusing on the FS around $\bar{K}$,
with decreasing the energy, the two contours ($\alpha$ and $\beta$) evolve into three sheets ($\alpha$, $\kappa$,
and $\gamma$), indicating that there are three bands around $\bar{K}$, and two of them ($\kappa$ and $\gamma$)
could be degenerate near $E_{\rm F}$. The ARPES spectra along the $\bar{\Gamma}$-$\bar{K}$-$\bar{M}$ and $\bar{\Gamma}$-$\bar{M}$ directions
are presented in \textcolor{black}{Figs. 1(g)-1(j)}.
We observe multiple linear dispersions around $\bar{K}$, which are denoted as $\alpha$, $\kappa$, and $\gamma$
as in Figs. 1(d)-1(f). The near-$E_{\rm F}$ intensity of $\kappa$ and $\gamma$ bands is weak in the first BZ, consistent with the mappings in Figs. 1(d) and 1(e). By extrapolating the branch of $\kappa$ band at higher binding energies and tracing the $\gamma$ band in the second BZ, one expects three Dirac-like crossings around $\bar{K}$.
The $\alpha$ and $\kappa$ bands cross at $\sim$-0.20 eV, the $\kappa$ and $\gamma$ bands could be degenerate near $E_{\rm F}$, and the crossing
of two branches of $\gamma$ band locates at $\sim$-0.30 eV.

As suggested by the calculations in CDW phase, the band structure is strongly modified around $\bar{K}$ as compared to that in pristine phase \cite{TanHX2021}. The most prominent reconstruction
is a large gap opening at the Dirac-like crossing above $E_{\rm F}$ along the $\bar{\Gamma}$-$\bar{K}$ direction. Here, the crossing corresponds to the one between the $\kappa$ and $\gamma$ bands.
We record the intensity in the second BZ to study the CDW-related
gap opening along the $\bar{\Gamma}$-$\bar{K}$ direction. As shown in \textcolor{black}{Fig. 2(a)}, we reveal that the $\kappa$ and $\gamma$
bands cross below $E_{\rm F}$ and there is an almost flat feature $\beta$ close to $E_{\rm F}$ separated by depletion of intensity, pointing to a
possible gap opening. In the corresponding energy distribution curves (EDCs) \textcolor{black}{[Fig. 2(b)]}, as indicated by the black curve, a
distinct energy gap of $\sim$30 meV is resolved between the $\beta$ band ($\sim$0.02 eV below $E_{\rm F}$) and the $\kappa$/$\gamma$ band, where the
degeneracy occurs at $\sim$0.05 eV below $E_{\rm F}$. Intriguingly, we observe a peak-dip-hump feature around the energy gap, the lineshape is
composed of the dispersionless $\beta$ band close to $E_{\rm F}$ (peak) and the linearly dispersive $\kappa$/$\gamma$ band at higher binding
energy (hump). \textcolor{black}{Similar peak-dip-hump structure has also been reproduced in a different sample (Fig. S2).}

\textcolor{black}{Figures 2(e)-2(g) show the electronic structure in pristine phase on a sample cleaved at 150 K (\#S2). The visible features at $E_{\rm F}$
[Fig. 2(e)] are analogous to that in Fig. 1(d), implying the weak CDW modulation potential. The peak-dip-hump structure disappears around the crossing of $\kappa$* and $\gamma$* bands [Fig. 2(g)], further demonstrating it relates to the CDW formation (see Fig. S3 for additional temperature dependent data of the peak-dip-hump feature).}
To determine how the CDW gap and peak-dip-hump structure distribute in momentum space, we perform measurements along cut \#2 at $T$ = 1.4 K,
illustrated by the white arrow in the inset of \textcolor{black}{Fig. 2(c)}. The gap size and peak-dip-hump lineshape exhibit little change
\textcolor{black}{[Fig. 2(d)]}. \textcolor{black}{Similar behaviors are also observed in the spectra along cuts \#3 and \#4 around $\bar{K}$
(Fig. S4). Consequently, these CDW-related features are uniformly distributed on the $\beta$ FS.}

\begin{figure}[t]
  \vspace{-0.63cm}
  \setlength{\abovecaptionskip}{-0.2cm}
  \setlength{\belowcaptionskip}{-0.15cm}
  \begin{center}
  \includegraphics[trim = 4.5mm 0mm 0mm 0mm, clip=true, width=1.04\columnwidth]{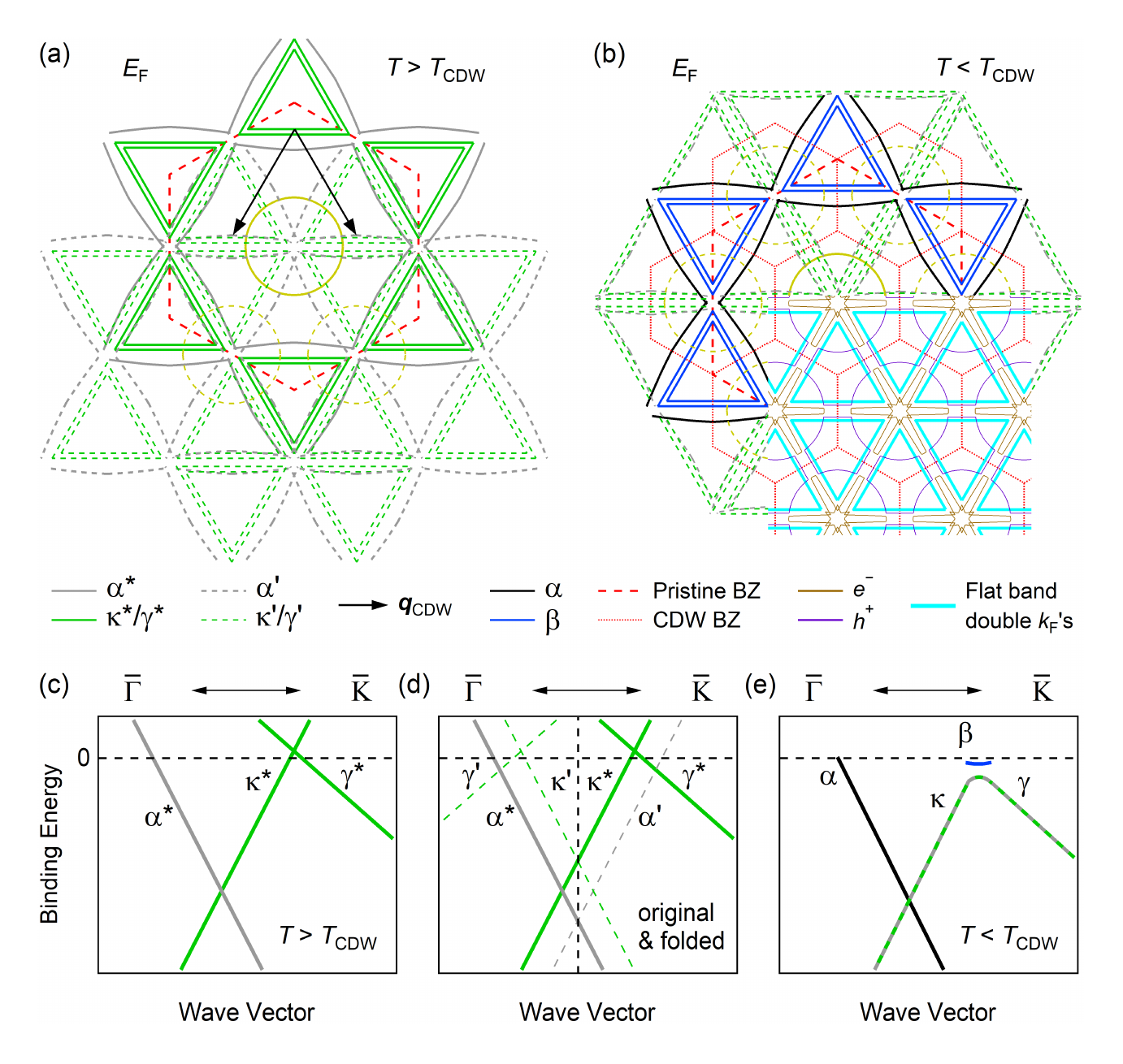}
  \end{center}
  \caption{
  \textcolor{black}{(a),(b) Sketches of the experimental FSs (solid curves) above and below $T_{\rm CDW}$ together with the folded ones (dashed curves) by the
  2 $\times$ 2 CDW vector, respectively.
  The inset of (b) illustrates the reconstructed FSs in CDW BZs. The cyan triangles are bolded to indicate the nature of double-$k_{\rm F}$ crossings
  of flat bands.
  (c)-(e) Band structure cartoon along the $\bar{\Gamma}$-$\bar{K}$ direction above [(c),(d)] and below [(e)] $T_{\rm CDW}$, respectively.
  The $\alpha$*, $\kappa$*, and $\gamma$* bands represent the original band structure in pristine phase, and the $\alpha$', $\kappa$', and $\gamma$' bands
  are the corresponding folded band structure with respect to the momentum cut, where the vertical dashed line locates in (d), connecting adjacent $\bar{M}$ points.}
  }
\end{figure}

\textcolor{black}{Recent calculations suggest that the electronic instability via FS nesting is important in the CDW formation of $A$V$_3$Sb$_5$ \cite{TanHX2021,ChoS2021}.}
From a Peierls perspective, the electronic susceptibility would develop a logarithmic divergence at some well nested FS sheets, which could
drive a CDW transition \cite{Gruner1994,XTZhu2017}.
To examine the nesting property of overall FS, we sketch the FS above $T_{\rm CDW}$ and fold it by two representative
$\emph{\textbf{q}}_{\rm CDW}$ = ($\pi$, 0) as shown in \textcolor{black}{Fig. 3(a), where the $\kappa$*/$\gamma$* sheets are doubled to indicate the nature of two-$k_{\rm F}$ crossings [see more discussion in Sec. 6 of Supplemental Material (SM)].} One obtains that nearly perfect nestings exist between the original $\gamma$* \textcolor{black}{(inner solid triangle)} and folded $\alpha$' as well as the original $\alpha$* and folded $\gamma$' \textcolor{black}{(inner dashed  triangle)}. \textcolor{black}{Upon entering the CDW phase, the interactions involved in nested bands and overlaps could result in gap opening and FS reconstructions in the CDW BZ as schematically shown in the inset of Fig. 3(b), where the hybridizations between cyan and purple contours are omitted for
better visualization of flat band. However, since the modulation potential of CDW is weak in CsV$_3$Sb$_5$
\cite{WangZX2021}, some FS contours could be strongly suppressed, then such reconstructions in the CDW BZ are not detected in ARPES. The observations in Fig. 1(d) do not represent the complete FS in CDW state.}

\begin{figure}[t]
  \vspace{-0.63cm}
  \setlength{\abovecaptionskip}{-0.1cm}
  \setlength{\belowcaptionskip}{-0.15cm}
  \begin{center}
  \includegraphics[trim = 4.5mm 0mm 0mm 0mm, clip=true, width=1.02\columnwidth]{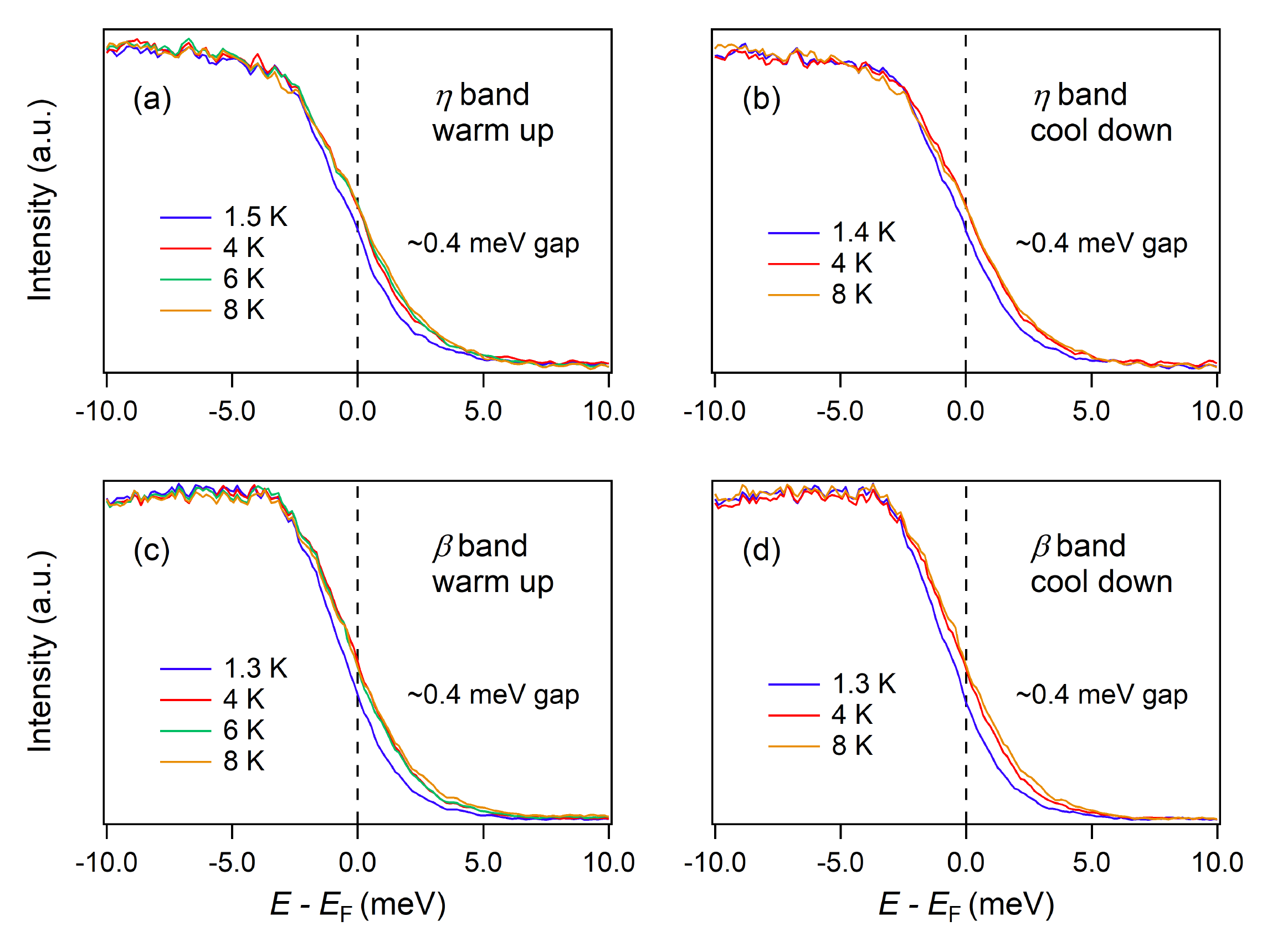}
  \end{center}
  \caption{
  \textcolor{black}{(a),(c) Temperature dependent EDCs taken at $k_{\rm F}$'s of the $\eta$ and $\beta$ bands along the $\bar{\Gamma}$-$\bar{K}$ and $\bar{\Gamma}$-$\bar{K}$-$\bar{M}$ directions (pristine BZ), where the sample is warmed up from 1.5 and 1.3 K to 8 K, respectively.
  (b),(d) Same as (a),(c) by cooling down the sample from 8 K to 1.4 and 1.3 K, respectively.}
  }
\end{figure}

Since the band folding by $\emph{\textbf{q}}_{\rm CDW}$ is equivalent to that along the $\bar{\Gamma}$-$\bar{K}$ direction with respect to the momentum cut connecting neighboring $\bar{M}$ points, we show the band structure cartoons along the $\bar{\Gamma}$-$\bar{K}$ direction to clarify the \textcolor{black}{possible}  mechanism.
We sketch the original bands ($\alpha$*, $\kappa$*, and $\gamma$*) above $T_{\rm CDW}$ in \textcolor{black}{Fig. 3(c)} and plot them in \textcolor{black}{Fig. 3(d)} together with
the folded bands ($\alpha$', $\kappa$', and $\gamma$') expected to emerge due to CDW. It is seen that $\alpha$' will interact with $\gamma$* just below
$E_{\rm F}$, resulting in the CDW gap opening which leads to the flat band $\beta$ and peak-dip-hump lineshape. The $\kappa$* band does not
have a ``partner" to interact with in the CDW state and therefore remains as is, smearing out the peak-dip-hump lineshape at lower momentum
values. Thus, the $\kappa$ feature in \textcolor{black}{Fig. 3(e)} actually represents a superposition of $\kappa$* and $\alpha$'. Analogous peak-dip-hump
structure is not resolved at equivalent momentum [$\sim$0.4 \AA$^{-1}$ in \textcolor{black}{Figs. 2(a) and 2(c)}] where $\alpha$* should interact with $\gamma$',
\textcolor{black}{because the latter, being folded of $\gamma$*, is vanishingly weak compared with the former and the gapped structure is hidden by the high-temperature intensity distribution (see more discussion in Sec. 7 of SM).} Most of the folded bands are hardly observed in CsV$_3$Sb$_5$ because strongest features (e.g. $\alpha$') nearly overlap with already existing ones (e.g. $\kappa$*) and the modulation potential appears to be weak \cite{WangZX2021}. The formation of CDW gap can explain the band reconstructions reported in a recent ARPES study \cite{HuY2021}. The gap size of $\sim$30 meV is comparable with that in previous STM \cite{JiangY2020} and ARPES \cite{KangM2021} studies.
Note that theory predicts also a spin-orbit coupling
(SOC) induced gap at the crossing of $\kappa$* and $\gamma$* in the pristine phase \cite{TanHX2021}.
As schematically illustrated in \textcolor{black}{Fig. S5}, the $\beta$ feature appears to be flattened by both gaps -- from SOC gap above and from CDW gap below. However, since weaker features are not clearly
resolved in our spectra, we leave only most pronounced of them in \textcolor{black}{Fig. 3(e)}. \textcolor{black}{Our data, together with earlier observations \cite{JiangY2020,KangM2021}, suggest that the theory may overestimate the size of CDW gap (see Fig. S6 and Sec. 9 of SM for more discussion on the correspondence and difference between experiments and calculations).}

\textcolor{black}{Last but not least, we further make an attempt to detect the superconducting gap on the multiple FSs around $\bar{\Gamma}$ and $\bar{K}$
(pristine BZ). As presented in Fig. 4(a), on the $\eta$ band at $\bar{\Gamma}$, when we increase the temperature from 1.5 K to 4 K, a leading edge shift or leading edge gap (LEG) of $\sim$0.4 meV is revealed. After warming up the sample to 8 K, one obtains that the leading edge midpoints (LEM) of the EDCs in the non-superconducting state coincide and locate at $E_{\rm F}$. Then we cool down the sample back to 1.4 K, the LEM of the spectra at 8 K and 4 K still coincide and again there is a LEG of $\sim$0.4 meV at 1.4 K [Fig. 4(b)]. These facts can rule out a rigid band shift along with the temperature and demonstrate the superconducting gap opening. We perform similar measurements on the flat $\beta$ band, the LEG of $\sim$0.4 meV is also observed [Figs. 4(c) and 4(d)], implying the multiband superconductivity in
CsV$_3$Sb$_5$. The values of $\Delta$ $\sim$ 0.4 meV and $T_{\rm c}$ $\sim$ 2.5 K would yield 2$\Delta$/$k_{\rm B}$$T_{\rm c}$ $\approx$ 3.7. Although the LEG is a good qualitative measure of the superconducting gap \cite{ZXShen1993}, its quantitative correspondence with the real gap is more complicated and depends on many factors \cite{AAKordyuk2003}.
As a rule, the real gap is slightly larger than the LEG.}

\textcolor{black}{The competing relationship between CDW and superconductivity is uncovered in recent pressure and strain measurements on CsV$_3$Sb$_5$
\cite{YuFH2021NC,ChenKY2021,QianTM2021}.
In a Bilbro-McMillan partial gaping scenario, the CDW would strongly compete
with superconductivity for the DOS at $E_{\rm F}$ \cite{BilbroG1976}. Such interplay resembles many other systems with intertwined CDW and superconductivity, like Cu$_x$TiSe$_2$ \cite{MorosanE2006}, pressurized 1$T$-TiSe$_2$ \cite{KusmartsevaAF2009}, Pd-intercalated rare-earth tritellurides \cite{HeJB2016}, and cuprates \cite{ChangJ2012}. However, we find that the DOS at $E_{\rm F}$ over multiple FSs around $\bar{\Gamma}$ and $\bar{K}$ is not depleted by the CDW here. There is no CDW gap opening on the $\eta$ (Fig. S7) and $\alpha$ [Figs. 2(a)-2(d) and S4] FSs, the DOS of $\beta$ FS
could be enhanced by the CDW-induced flat band (see more detailed discussion in Sec. 9 of SM).
Regrading the reconstructed FS in CDW state [inset of Fig. 3(b)], which actually determines the underlying physics at low temperatures, besides the $\beta$ FS, the $\alpha$ FS is also mostly defined by the flat band. The superconducting gap is observed on both the cyan and purple contours in CDW BZ. As is the case for most known superconductors, the near-$E_{\rm F}$ DOS boosted by a saddle point or an extended flat region of the dispersion could play an important role in the superconductivity \cite{Borisenko2010,Borisenko2013}. Consequently, the available electronic states at $E_{\rm F}$ in the CDW phase are most likely in favor of the multiband superconductivity, particularly the enhanced DOS associated with the flat band, which is estimated to have an extension of $\sim$54.2\% of the total perimeter of FS in CDW BZ.
The occurrence of the extended flat feature is also energetically favorable as illustrated in Fig. S8. The actual superconducting gap value on the flat band is most likely larger than 0.4 meV, because the $\beta$ is involved in a peak-dip-hump structure, where gapping the peak may cause the transfer of some spectral weight to the dip and the resulting LEG could be reduced. Therefore, although the CDW and superconductivity compete for the similar electronic states at $E_{\rm F}$, the formation of CDW seems to prepare the seedbed for the multiband superconductivity at lower temperatures in CsV$_3$Sb$_5$.}

\textcolor{black}{In summary, we have studied the CDW and superconducting states of CsV$_3$Sb$_5$ by ARPES with high resolution.
We observe a CDW-induced peak-dip-hump lineshape around $\bar{K}$, demonstrating the presence of an isotropic CDW gap and a flat band close to
$E_{\rm F}$. The FS nesting is suggested to play a role in driving CDW instabilities. The superconducting gap of $\sim$0.4 meV is detected over multiple FSs around $\bar{\Gamma}$ and $\bar{K}$, implying multiband superconductivity. The finite DOS at $E_{\rm F}$ in the CDW phase is most likely in favor of the  superconductivity, particularly the boosted DOS associated with the flat band. Our results facilitate further solving the controversial origin of CDW and
understanding its interplay with superconductivity in CsV$_3$Sb$_5$.}

\begin{acknowledgments}
This work was in part supported by the Deutsche Forschungsgemeinschaft under Grants SFB 1143 (project C04). R.L. was supported by the National
Natural Science Foundation of China (Grant No. 11904144). A.K., B.B., and S.B. acknowledge the support from BMBF via project UKRATOP. H.C.L.
was supported by the National Natural Science Foundation of China (Grants No. 11822412 and No. 11774423), the Ministry of Science and Technology of
China (Grants No. 2018YFE0202600 and No. 2016YFA0300504), and the Beijing Natural Science Foundation (Grant No. Z200005). A.F., E.S., B.B., and S.B.
acknowledge the support from W{\"u}rzburg-Dresden Cluster of Excellence on Complexity and Topology in Quantum Matter--$\emph{ct.qmat}$ (EXC 2147, project-id 390858490).

R.L., A.F., and Q.W.Y. contributed equally to this work.
\end{acknowledgments}

\end{document}